\documentclass[adraft,copyright,creativecommons]{eptcs}
\providecommand{\event}{October 2020\\Revised December 2020} 
\usepackage{breakurl}                 
\usepackage{underscore}               

\usepackage{amsfonts,csquotes,scalerel,tikz-cd,adjustbox,float,subcaption,pgfplots}
\usetikzlibrary{shapes.misc}
\tikzset{gp2 node/.style={draw, circle, thick, minimum width=0.64cm}}
\tikzset{root node/.style={draw, circle, thick, minimum width=0.64cm, double, double distance=0.3mm}}
\pgfplotsset{compat=1.15}

\newenvironment{allintypewriter}{\ttfamily}{\par}

\definecolor{gp2green}{RGB}{69, 191, 156}
\definecolor{gp2blue}{RGB}{153, 187, 255}
\definecolor{gp2red}{RGB}{236, 107, 116}
\definecolor{gp2pink}{RGB}{239, 161, 193}
\definecolor{gp2grey}{RGB}{196, 192, 200}

\definecolor{performanceBlue}{RGB}{0, 136, 255}
\definecolor{performanceYellow}{RGB}{252, 199, 17}

\title{The Improved GP\,2 Compiler}
\author{
Graham Campbell\thanks{Supported by a Vacation Internship and a Doctoral Training Grant No. (2281162) from the Engineering and Physical Sciences Research Council (EPSRC) in the UK, while at University of York and Newcastle University, respectively.}\institute{School of Mathematics, Statistics and Physics, Newcastle University\\Newcastle upon Tyne, United Kingdom}\email{g.j.campbell2@newcastle.ac.uk}
\and
Jack Rom\"o\institute{Mathematical Institute, University of Oxford, Oxford, United Kingdom}\email{jack.romo@lincoln.ox.ac.uk}
\and
Detlef Plump\institute{Department of Computer Science, University of York, York, United Kingdom}\email{detlef.plump@york.ac.uk}
}
\def\titlerunning{The Improved GP\,2 Compiler}
\def\authorrunning{G. Campbell, J. Rom\"o, and D. Plump}
\begin{document}
\maketitle

\begin{abstract}
GP\,2 is a rule-based programming language based on graph transformation rules which aims to facilitate program analysis and verification. Writing efficient programs in such a language is challenging because graph matching is expensive. GP\,2 addresses this problem by providing rooted rules which, under mild conditions, can be matched in constant time. Recently, we implemented a number of changes to Bak's GP\,2-to-C compiler in order to speed up graph programs. One key improvement is a new data structure for dynamic arrays called \texttt{BigArray}. This is an array of pointers to arrays of entries, successively doubling in size. To demonstrate the speed-up achievable with the new implementation, we present a reduction program for recognising binary DAGs which previously ran in quadratic time but now runs in linear time when compiled with the new compiler.
\end{abstract}

\section{Introduction} \label{chapter:intro}

Rule-based graph transformation has been a topic of research since the early 1970s and has been the subject of numerous articles of both a theoretical and practical nature (see the monographs \cite{Ehrig-Ehrig-Prange-Taentzer06a,Ehrig-Ermel-Golas-Hermann15a}). In addition, various graph programming languages and model transformation tools have been developed, including AGG \cite{Runge-Ermel-Taentzer11a}, GReAT \cite{Agrawal-Karsai-Neema-Shi-Vizhanyo06a}, GROOVE \cite{Ghamarian-Mol-Rensink-Zambon-Zimakova12a}, GrGen.Net \cite{Jakumeit-Buchwald-Kroll10a}, Henshin \cite{Arendt-Biermann-Jurack-Krause-Taentzer10a} and PORGY \cite{Fernandez-Kirchner-Mackie-Pinaud14a}. This paper focuses on the implementation of GP\,2, an experimental programming language based on graph transformation rules \cite{Plump12a}. GP\,2 has been designed to support formal reasoning on programs \cite{Poskitt-Plump12a,Plump16a} and comes with a formal semantics in the style of structural operational semantics.

GP\,2 programs manipulate directed graphs whose nodes and edges are labelled with heterogeneous lists of integers and character strings. The principal programming construct in GP\,2 are conditional graph transformation rules labelled with expressions. Rules operate on host graphs according to the attributed graph transformation framework of \cite{Hristakiev-Plump16a}: in a two-stage process, the rule is first instantiated by replacing all variables with values of the same type and evaluating all expressions, yielding a rule in the double-pushout approach with relabelling \cite{Habel-Plump02a}. In the second stage, the instantiated rule is applied to the host graph by constructing two suitable pushouts. 

The performance bottleneck for GP\,2 (and graph transformation in general) is matching the left-hand graph $L$ of a rule within a host graph $G$, requiring time polynomial in the size of $L$ \cite{Bak-Plump12a}. As a consequence, linear-time graph algorithms in imperative languages may be slowed down to polynomial time when they are recast as rule-based programs. To speed up matching, GP\,2 supports \emph{rooted} graph transformation where graphs in rules and host graphs are equipped with so-called root nodes, originally developed by D\"orr \cite{Dorr95a}. Roots in rules must match roots in the host graph so that matches are restricted to the neighbourhood of the host graph's roots.

Using GP\,2 programs with rooted rules, it has been possible to solve the 2-colouring problem for graphs of bounded degree in linear time \cite{Bak-Plump16a}. Remarkably, the program generated by Bak's GP\,2-to-C compiler matches the running time of Sedgewick's hand-crafted C implementation of the 2-colouring problem \cite{Sedgewick02a}. Meanwhile a few more conventional linear-time graph algorithms have been shown to have GP\,2 implementations running in linear time on bounded-degree graphs \cite{Campbell-Courtehoute-Plump19b}. To the best of our knowledge, no comparable performance results for low complexity graph algorithms have been reported for any other graph transformation language. 

In this paper, we outline some performance issues of the existing GP\,2-to-C compiler and the changes we made to address them, motivating each change, with examples where relevant. We also explain why we have not reached for off the shelf garbage collection algorithms. The new compiler produces programs with asymptotic runtime performance either the same as the original, or strictly faster. We give some timing results measuring the performance of the output programs of the original and improved compiler, providing empirical evidence both of our improvements and lack of performance degradation in cases beyond the scope of the improvements. We also implemented a root-reflecting mode to the compiler to allow a clean theoretical treatment of rooted rules.

\section{Bak's GP\,2 Compiler} \label{section:bakcompiler}

Before we discuss our modifications to the GP\,2 Compiler\footnote{The new compiler is available at \url{https://github.com/UoYCS-plasma/GP2}}, we first outline its prior state. The compiler detailed in Bak's thesis \cite{Bak15a} compiled GP\,2 programs into C code, which was then compiled by the GCC compiler into an executable. The original compiler stored a graph as two dynamic arrays, one for nodes and one for edges. Internally, each of these contained two arrays, one of the actual elements and another of indices that were empty, or \enquote{holes}. In the case of nodes, we dub these the node and hole arrays respectively. When iterating through nodes, each index would have to be checked to ensure it was not a hole. Deleting a node would require tracking the new hole and inserting nodes could be done by filling a hole should one exist. This raised performance issues if a program deleted a large number of nodes, for instance, in a graph reduction algorithm, as the enormous number of holes would make traversing the final smaller graph as slow as the original larger one. 

\begin{figure}[H]
\centering
{
\begin{subfigure}{.48\textwidth}
\centering
\pgfplotsset{scaled y ticks=false}
\begin{tikzpicture}[scale=0.7]
\begin{axis}[
xlabel={Number of nodes in input graph},
ylabel={Execution time (ms)},
xmin=0, xmax=1100000,
ymin=0, ymax=1000,
legend pos=north west,
ymajorgrids=true,
grid style=dashed,
yticklabel style={/pgf/number format/fixed},
]
\addplot[color=performanceBlue, mark=square*] 
coordinates {
    (100000,114.80)
    (200000,199.31)
    (300000,280.68)
    (400000,367.15)
    (500000,457.26)
    (600000,545.06)
    (700000,636.08)
    (800000,730.74)
    (900000,826.28)
    (1000000,921.32)
};
\addlegendentry{New Impl.}
\end{axis}
\end{tikzpicture}
\end{subfigure}
\begin{subfigure}{.48\textwidth}
\centering
\vspace{0.5em}
\pgfplotsset{scaled y ticks=false}
\begin{tikzpicture}[scale=0.7]
\begin{axis}[
xlabel={Number of nodes in input graph},
ylabel={Execution time (ms)},
xmin=0, xmax=110000,
ymin=0, ymax=16000,
legend pos=north west,
ymajorgrids=true,
grid style=dashed,
yticklabel style={/pgf/number format/fixed},
]
\addplot[color=performanceYellow, mark=square*] 
coordinates {
    (10000,128.14)
    (20000,460.44)
    (30000,1017.36)
    (40000,1806.24)
    (50000,2981.68)
    (60000,4460.76)
    (70000,6321.76)
    (80000,8577.40)
    (90000,11436.25)
    (100000,14617.81)
};
\addplot[color=performanceBlue, mark=square*] 
coordinates {
    (10000,20.17)
    (20000,28.81)
    (30000,38.40)
    (40000,44.79)
    (50000,54.45)
    (60000,62.15)
    (70000,70.16)
    (80000,77.73)
    (90000,86.29)
    (100000,94.46)
};
\addlegendentry{Old Impl.}
\addlegendentry{New Impl.}
\end{axis}
\end{tikzpicture}
\end{subfigure}
}
\vspace{-0.4em}
\caption{Measured performance of \texttt{is-discrete.gp2}}
\label{fig:is-discrete-timing}
\end{figure}
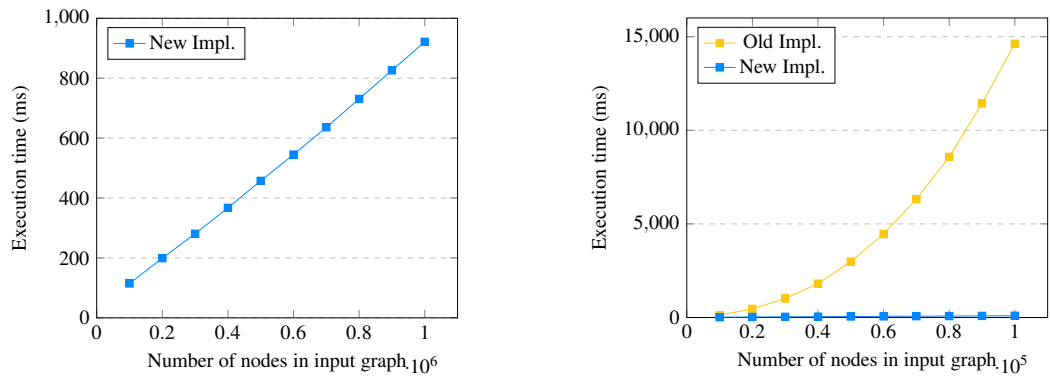

We demonstrate this effect by the toy reduction program in Figure \ref{fig:is-discrete-gp2} which recognises edgeless graphs. (The loop \texttt{del!} applies the rule \texttt{del} as long as possible; the input graph is edgeless if and only if the program does not fail.) This program should run in linear time, as finding an arbitrary node should take constant time. Alas, each deleted node adds a new hole at the start of the node array, making the program take quadratic time due to traversing the holes at each rule match. The measured performance of both the original and improved implementation on discrete graphs is provided in Figure \ref{fig:is-discrete-timing}.

\begin{figure}[H]
\centering
\fbox{\begin{minipage}{.8\textwidth}
\begin{allintypewriter}
Main = del!; if node then fail

\setlength{\tabcolsep}{0.4cm}

\medskip
\smallskip

\begin{tabular}{ p{4.68cm} p{4.68cm} }
	
	del(x:list) & \vspace{-2mm} node(x:list) \\
	
	\begin{tikzpicture}
		\node (a) at (0,0)       [gp2 node] {x};
		
		\node (b) at (.875,0)    {$\Rightarrow$};
		
		\node (c) at (1.75,0.02) {$\emptyset$};
		
		\node (A) at (0,-.45)    {\tiny{\,}};
	\end{tikzpicture}
	&
	
	\vspace{-9mm}
	\begin{tikzpicture}
		\node (a) at (0,0)       [gp2 node] {x};
		
		\node (b) at (.875,0)    {$\Rightarrow$};
		
		\node (c) at (1.75,0)    [gp2 node] {x};
		
		\node (A) at (0,-.45)    {\tiny{1}};
		\node (C) at (1.75,-.45) {\tiny{1}};
	\end{tikzpicture}
	\\
\end{tabular}
\end{allintypewriter}
\end{minipage}}
\vspace{-0.4em}
\caption{The program \texttt{is-discrete.gp2}}
\label{fig:is-discrete-gp2}
\end{figure}

Should the node array be too small for a new entry, it would be doubled with the \texttt{realloc()} C standard library function, the same being done for the hole array. However, this could change the array's position in memory, making any pointers to nodes invalid. This meant that Bak needed to store indices of nodes instead of pointers, adding extra memory operations to resolve indices every time a node was accessed.

For root nodes, Bak added a linked list to each graph, each entry holding a pointer to a root node in the graph's node array. Nodes would contain a flag themselves detailing if they were a root node or not. Iterating through or deleting root nodes would now take constant time, should there be a constant upper bound on the number of root nodes.

A node itself contained its own index, the number of edges for which it is a source or target, termed its outdegree and indegree respectively, its incident edges, label, mark and several boolean flags. The incident edges to a node were stored in a unique manner, with indices of the first two edges being stored statically as part of the node type itself, and the rest in two dynamically allocated arrays of incoming and outgoing edges. This would avoid allocating arrays for a node should its degree be below three. Each edge would contain its own index, label, mark, and indices of its source and target nodes. It would, similarly to nodes, hold a boolean flag of whether it had been matched or not yet, to prevent overlapping matches.

Bak's compiler implementation included a parser for input graphs, necessary to allow programs to accept user provided input graphs. Unfortunately this implementation could not dynamically allocate memory to accommodate arbitrarily large input graphs. That is, only small inputs could be correctly processed, with memory overflows not gracefully handled or even necessarily detected.

Finally, to accommodate programs running within the condition of an \texttt{if} statement, for instance, a stack of states was needed. Bak implemented graph stacks in two varieties: copying the previous graph into a stack to reuse it when unwinding, and storing changes to the graph in the stack to undo them in reverse when unwinding. The former simply stored all the data that a node or edge contained, reconstructing the node or edge as needed as it unwound. Graph copying would simply perform a deep copy of the graph as expected.

\section{The Improved Compiler} \label{section:improvedimpl}

We have updated the compiler implementation to address the issues described in Section \ref{section:bakcompiler}. We now present our improvements, which were first documented in the unpublished report \cite{Campbell-Romo-Plump19a}. We draw root nodes using double circles.

\subsection{Graph Parsing}

To resolve the issues with parsing, we decided to employ Judy arrays\footnote{\url{http://judy.sourceforge.net}} \cite{Silverstein02a}, instead of a simple dynamic array. Invented by Doug Baskins, Judy arrays are a highly cache-optimised hash table implementation. The size of a Judy array is not statically pre-determined but is adjusted, at runtime, to accommodate the number of keys, which themselves can be integers or strings. Instead of storing nodes in the array directly, we now store pointers to nodes in the host graph as Judy arrays can only store references to a single word of data. Reallocating the array when doubling it could move the array around and invalidate previous pointers, an issue we resolve in the next subsection. This allowed an edge to retrieve pointers to its source and target efficiently due to Judy arrays' fast runtime performance \cite{Luan-Du-Wang-Ni-Chen07a}. This also resolved problems with unnecessary node array size, allowing node IDs to be arbitrarily large without causing memory problems, as the array simply saw these IDs as meaningless keys. We consider these improvements to be bug fixes, rather than performance improvements, unlike our other internal data structure changes.

\subsection{From Arrays to Linked Lists} \label{sec:arraystolists}

To resolve the problems hole arrays pose, we switched to a linked list pointing to nodes. This change let us run the discrete graph deletion program in linear time, skipping holes we would otherwise traverse. Nodes, edges and now linked lists could then be stored in arrays doubling in size when needed as before. We choose this data structure above others that may have faster random access time, such as binary trees, because our only use case is iterating through the entire list to match subgraphs or adding and deleting nodes and edges. Accessing an arbitrary element of the data structure quickly is therefore irrelevant, meaning linked lists suit us perfectly.

It then became apparent that node indices were redundant, adding unneeded memory operations to resolve a node's address. Replacing indices with direct pointers would in turn add the problem of pointers being invalidated should the array of nodes or edges be moved when \texttt{realloc()} is called to enlarge them. To fix this, we replaced all internal arrays with a new type we dubbed \texttt{BigArray}. The \texttt{BigArray} type is an array of pointers to arrays of entries, successively doubling in size. Accessing a given index is constant time, using the position of the largest set bit in the index to identify which sub-array to access. A logarithmic number of memory allocations are performed overall when filling the array with entries, with the array of arrays being reallocated $\mathrm{O}(\mathrm{log}(\mathrm{log}(n)))$ times. While such doubling arrays could lead to memory thrashing, in the case of GP\ 2 we are not interested in working with graphs of such a scale that this would typically happen.

Big arrays also manage holes like the prior implementation did. However, instead of using a second array of holes, big arrays store a linked list of holes within the hole entries in the array, keeping a pointer to the first hole. When a hole is created in the array, that position in the array is overwritten with the data of a new linked list entry, becoming the head of the list of holes. This avoids having to use extra memory for holes, making the \texttt{BigArray} type more memory efficient and making deletion of elements constant time.

Most importantly, big arrays allow one to allocate more memory to the array without having to possibly move previous entries in memory, simply creating a new array. This means the low number of memory allocations may be maintained without pointers to nodes and edges being invalidated. Thus, three big arrays are now stored within a graph, one for nodes, one for edges, and one for entries in the linked list of nodes, termed \texttt{NodeList}. A node list simply contains a pointer to the node it refers to and a pointer to the next entry in the linked list. The same is true of edge lists, albeit for edges.

Each node now contains a big array of linked list entries for edges and pointers to the linked list of outgoing edges and of incoming edges. No iteration through edges directly is ever needed beyond printing a graph, which can be done by iterating through the outgoing edges of every node, so no total list of edges is maintained. A list of edges is again acceptable: our only use case is iteration. Furthermore, as with the reasoning to add root nodes to GP\ 2, we are often interested in graphs of bounded degree, making fast random access of edges even less of a priority.

To avoid all pointers to a node or edge being garbage once it is deleted, nodes and edges were also made to track who references them, only being garbage collected when no such references exist. Nodes now hold flags representing if they are in a graph or referred to in the stack of graph changes, and edges remember if they are in a node's list of incoming/outgoing edges or in the stack also. Should a node or edge be deleted, the operation can be deferred should other references still exist. Now, stacks of graph changes can simply store pointers to nodes and edges rather than any data in them, as deletion would be automatic. We opted to use a more manual approach to garbage collection rather than existing implementations like the Boehm-Demers-Weiser garbage collector \cite{Boehm-Demers-WeiserGC}, as managing this ourselves massively reduces the amount of system calls needed and lets us even consider allocating memory in chunks doubling in size as we have done. Furthermore, garbage collection is not the core content of our improvements, as our focus remains on eliminating complexity issues in the aforementioned examples.

In the previous version of the compiler, there was only a single edge array associated with each node. Now each node has, in addition to an internal big array storing edges, two separate edge lists, for outgoing and incoming edges respectively. It is now possible to run programs that previously required bounded degree to obtain a certain worst case time complexity, now with only bounded incoming degree, or bounded outgoing degree, since search plans can now only consider edges of the correct orientation.

\subsection{Root-Reflecting Mode} \label{sec:rootrefl}

Finally, we have added a new \enquote{root-reflecting} mode to the compiler, allowing the programmer to decide if rule application should reflect and well as preserve root nodes. The motivation for this change is due to an issue in the theoretical foundation of rooted graph transformation with relabelling originally used by Bak and Plump which means that the right-hand square of a direct derivation need not be a natural pushout.

That is, in the usual definition of injective graph transformation, the application of a rule \(L \hookleftarrow K \hookrightarrow R\) to a graph \(G\), given a match \(L \to G\), is the computation of the pushout complement \((D, K \to D, D \hookrightarrow G)\) of \(L \to G\) and \(K \hookrightarrow L\), if it exists, and then the pushout \((H, D \hookrightarrow H, R \to H)\) of \(K \hookrightarrow R\) and \(K \to D\). The pushout complement exists exactly when the dangling condition is satisfied, and \(D\) and \(H\) are unique up to unique isomorphism. When Plump and Habel introduced relabelling in 2002 \cite{Habel-Plump02a}, in order to guarantee uniqueness they insisted on the pushout complement also being a pullback. This is no restriction, because both pushouts were already pullbacks in the original setting. By comparison, looking at Bak and Plump's explicit algorithm for computing derivations as the definition of rule application, we have two problems:

\begin{enumerate}
    \item The right square need not be a pullback.
    \item The left square need not be a pushout.
\end{enumerate}

\begin{figure}[H]
\centering
{
\begin{subfigure}{.48\textwidth}
\centering
\begin{tikzpicture}[every node/.style={inner sep=0pt, text width=6.5mm, align=center}]
    \node (a) at (0.0,0.0) [draw, circle, thick] {\,};
    \node (b) at (1.0,0.0) {$\leftarrow$};
    \node (c) at (2.0,0.0) [draw, circle, thick] {\,};
    \node (d) at (3.0,0.0) {$\rightarrow$};
    \node (e) at (4.0,0.0) [draw, circle, thick, double, double distance=0.4mm] {\,};

    \node (f) at (0.0,-1.0) {$\big\downarrow$};
    \node (g) at (1.0,-0.8) {\mbox{PB \checkmark}};
    \node (g) at (1.0,-1.2) {\mbox{PO \checkmark}};
    \node (h) at (2.0,-1.0) {$\big\downarrow$};
    \node (g) at (3.0,-0.8) {\mbox{PB $\times$}};
    \node (g) at (3.0,-1.2) {\mbox{PO \checkmark}};
    \node (j) at (4.0,-1.0) {$\big\downarrow$};

    \node (k) at (0.0,-2.0) [draw, circle, thick, double, double distance=0.4mm] {\,};
    \node (l) at (1.0,-2.0) {$\leftarrow$};
    \node (m) at (2.0,-2.0) [draw, circle, thick, double, double distance=0.4mm] {\,};
    \node (n) at (3.0,-2.0) {$\rightarrow$};
    \node (o) at (4.0,-2.0) [draw, circle, thick, double, double distance=0.4mm] {\,};
\end{tikzpicture}
\end{subfigure}
\begin{subfigure}{.48\textwidth}
\centering
\begin{tikzpicture}[every node/.style={inner sep=0pt, text width=6.5mm, align=center}]
    \node (a) at (0.0,0.0) [draw, circle, thick] {\,};
    \node (b) at (1.0,0.0) {$\leftarrow$};
    \node (c) at (2.0,0.0) {$\emptyset$};
    \node (d) at (3.0,0.0) {$\rightarrow$};
    \node (e) at (4.0,0.0) [draw, circle, thick] {\,};

    \node (f) at (0.0,-1.0) {$\big\downarrow$};
    \node (g) at (1.0,-0.8) {\mbox{PB \checkmark}};
    \node (g) at (1.0,-1.2) {\mbox{PO $\times$}};
    \node (h) at (2.0,-1.0) {$\big\downarrow$};
    \node (g) at (3.0,-0.8) {\mbox{PB \checkmark}};
    \node (g) at (3.0,-1.2) {\mbox{PO \checkmark}};
    \node (j) at (4.0,-1.0) {$\big\downarrow$};

    \node (k) at (0.0,-2.0) [draw, circle, thick, double, double distance=0.4mm] {\,};
    \node (l) at (1.0,-2.0) {$\leftarrow$};
    \node (m) at (2.0,-2.0) {$\emptyset$};
    \node (n) at (3.0,-2.0) {$\rightarrow$};
    \node (o) at (4.0,-2.0) [draw, circle, thick] {\,};
\end{tikzpicture}
\end{subfigure}
}
\caption{Problematic Derivation Examples}
\label{fig:root-refl}
\end{figure}
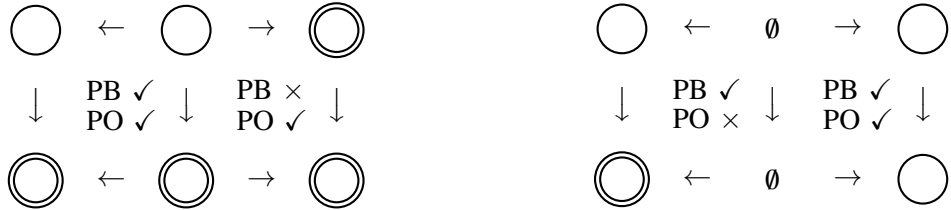

The first problem could be argued to not be major, since we remain compatible with the original notion of injective DPO graph transformation, and we still have uniqueness. However, there are two unfortunate consequences. The first is that derivations are no longer invertible in the sense that if \(G \Rightarrow_{r} H\), then \(H \Rightarrow_{r^{-1}} G\). The second is that non-root-reflecting matches may not properly capture the intention of a programmer. Take for example, the left diagram of Figure \ref{fig:root-refl}. The intention of a programmer writing such a rule would have been to match a non-root and make it rooted, but if morphisms are not required to reflect rootedness of nodes, the effect of applying the rule could be to do nothing, if the non-root was matched against a root. Requiring root-reflecting morphisms excludes this case, and the rule would, instead, be non-applicable.

The second problem is more serious, as it represents an irrevocable incompatibility with the definition of DPO graph transformation. This problem was discovered by Plump and Wulandari in 2020, and their example is shown in the right diagram of Figure \ref{fig:root-refl}.

Fixing both these problems (at the time, only aware of the first problem), Campbell developed a new foundation for rooted GT systems with relabelling \cite{Campbell19a}. Instead of only insisting on matches preserving root nodes, one must additional insist on them reflecting them too. This was formalised by defining rootedness using a partial function into a two-point set rather than pointing graphs with root nodes, thus allowing both squares in a derivation to be natural pushouts, where formally, rules are allowed to have undefined rootedness in their interface graphs. Plump and Wulandari also adopt Campbell's new formalism in their recent work \cite{Wulandari-Plump20a}.

\section{Timing Results} \label{section:timing}

We ran various benchmarks, comparing the old with the new implementation, some examples of which are included here. Our full range of experiments can be found in \cite{Campbell-Romo-Plump19a}, and the concrete syntax of the programs is also available\footnote{\url{https://gist.github.com/GrahamCampbell/c8d84d42e3913065d1f9859fd8aeb8dd}}.

In Section \ref{section:bakcompiler}, we observed that even the program that simply deleted all isolated nodes could not be executed in linear time by the old implementation. A more subtle example is when the program splits up the input graph as it executes. In Figure \ref{fig:bin-dag-recognition-program-det}, we give a rooted reduction program with this property that recognises binary directed acyclic graphs (DAGs).

\begin{figure}[H]
\centering
\noindent
\fbox{\begin{minipage}{.8\textwidth}
\begin{allintypewriter}
Main = (init; Reduce!; if flag then break)!; if flag then fail

Reduce = up!; try Delete else set\_flag

Delete = \{del1, del1\_d, del21, del21\_d, del22, del22\_d, del0\}

\medskip
\setlength{\tabcolsep}{16pt}
\vspace{0.3mm}
\begin{tabular}{  p{3.155cm}  p{6.2cm}  }
    
    \vspace{-1mm} init(x:list) & \vspace{-1mm} up(a,x,y:list) \\

    \vspace{-2mm}
    \adjustbox{valign=t}{\begin{tikzpicture}
        \node (a) at (0.0,0) [gp2 node] {x};

        \node (b) at (1.0,0) {$\Rightarrow$};
        
        \node (c) at (2.0,0) [root node] {x};
        
        \node (A) at (0.0,-.45) {\tiny{1}};
        \node (C) at (2.0,-.45) {\tiny{1}};
    \end{tikzpicture}}

    &

    \vspace{-2mm}
    \adjustbox{valign=t}{\begin{tikzpicture}
        \node (a) at (0.0,0) [root node] {x};
        \node (b) at (1.5,0) [gp2 node] {y};
        
        \node (c) at (2.5,0) {$\Rightarrow$};
        
        \node (d) at (3.5,0) [gp2 node] {x};
        \node (e) at (5.0,0) [root node] {y};
        
        \node (A) at (0.0,-.45) {\tiny{1}};
        \node (B) at (1.5,-.45) {\tiny{2}};
        \node (D) at (3.5,-.45) {\tiny{1}};
        \node (E) at (5.0,-.45) {\tiny{2}};
        
        \draw (b) edge[->,thick] node[above] {a} (a)
              (e) edge[->,thick,dashed] node[above] {a} (d);
    \end{tikzpicture}}
    \\
\end{tabular}
\begin{tabular}{  p{4.7cm}  p{4.6cm}  }

    \vspace{-1mm} del1(a,x,y:list) & \vspace{-1mm} del1\_d(a,x,y:list) \\

    \vspace{-2mm}
    \adjustbox{valign=t}{\begin{tikzpicture}
        \node (a) at (0.0,0)     [gp2 node] {x};
        \node (b) at (1.5,0)     [root node] {y};

        \node (c) at (2.5,0)     {$\Rightarrow$};

        \node (d) at (3.5,0)     [root node] {x};

        \node (A) at (0.0,-.45) {\tiny{1}};
        \node (D) at (3.5,-.45) {\tiny{1}};

        \draw (b) edge[->,thick] node[above] {a} (a);
    \end{tikzpicture}}

    &

    \vspace{-2mm}
    \adjustbox{valign=t}{\begin{tikzpicture}
        \node (a) at (0.0,0)     [gp2 node] {x};
        \node (b) at (1.5,0)     [root node] {y};

        \node (c) at (2.5,0)     {$\Rightarrow$};

        \node (d) at (3.5,0)     [root node] {x};

        \node (A) at (0.0,-.45) {\tiny{1}};
        \node (D) at (3.5,-.45) {\tiny{1}};

        \draw (b) edge[->,thick,dashed] node[above] {a} (a);
    \end{tikzpicture}}
    \\
\end{tabular}
\begin{tabular}{  p{4.7cm}  p{4.6cm}  }

    \vspace{-1mm} del21(a,b,x,y:list) & \vspace{-1mm} del21\_d(a,b,x,y:list) \\

    \vspace{-3mm}
    \adjustbox{valign=t}{\begin{tikzpicture}
        \node (a) at (0.0,0)     [gp2 node] {x};
        \node (b) at (1.5,0)     [root node] {y};

        \node (c) at (2.5,0)     {$\Rightarrow$};

        \node (d) at (3.5,0)     [root node] {x};

        \node (A) at (0.0,-.45) {\tiny{1}};
        \node (C) at (3.5,-.45) {\tiny{1}};

        \draw (b) edge[->,thick,bend left=-25] node[above] {a} (a)
              (b) edge[->,thick,bend left=25] node[above] {b} (a);
    \end{tikzpicture}}

    &

    \vspace{-3mm}
    \adjustbox{valign=t}{\begin{tikzpicture}
        \node (a) at (0.0,0)     [gp2 node] {x};
        \node (b) at (1.5,0)     [root node] {y};

        \node (c) at (2.5,0)     {$\Rightarrow$};

        \node (d) at (3.5,0)     [root node] {x};

        \node (A) at (0.0,-.45) {\tiny{1}};
        \node (C) at (3.5,-.45) {\tiny{1}};

        \draw (b) edge[->,thick,dashed,bend left=-25] node[above] {a} (a)
              (b) edge[->,thick,bend left=25] node[above] {b} (a);
    \end{tikzpicture}}
    \\
\end{tabular}
\begin{tabular}{ p{4.7cm}  p{4.6cm}  }

    \vspace{-1mm} del22(a,b,x,y,z:list) & \vspace{-1mm} del22\_d(a,b,x,y,z:list) \\

    \vspace{-2mm}
    \adjustbox{valign=t}{\begin{tikzpicture}
        \node (a) at (0.0,-0.0)  [gp2 node] {x};
        \node (b) at (0.0,-1.2)  [gp2 node] {y};
        \node (c) at (1.5,-0.6)  [root node] {z};

        \node (d) at (2.5,-0.6)  {$\Rightarrow$};

        \node (e) at (3.5,-0.0)  [root node] {x};
        \node (f) at (3.5,-1.2)  [gp2 node] {y};

        \node (A) at (0.0,-0.45) {\tiny{1}};
        \node (B) at (0.0,-1.65) {\tiny{2}};
        \node (E) at (3.5,-0.45) {\tiny{1}};
        \node (F) at (3.5,-1.65) {\tiny{2}};

        \draw (c) edge[->,thick] node[above, yshift=1pt] {a} (a)
              (c) edge[->,thick] node[above, yshift=1pt] {b} (b);
    \end{tikzpicture}}

    &

    \vspace{-2mm}
    \adjustbox{valign=t}{\begin{tikzpicture}
        \node (a) at (0.0,-0.00) [gp2 node] {x};
        \node (b) at (0.0,-1.2)  [gp2 node] {y};
        \node (c) at (1.5,-0.6)  [root node] {z};

        \node (d) at (2.5,-0.6)  {$\Rightarrow$};

        \node (e) at (3.5,-0.0)  [root node] {x};
        \node (f) at (3.5,-1.2)  [gp2 node] {y};

        \node (A) at (0.0,-0.45) {\tiny{1}};
        \node (B) at (0.0,-1.65) {\tiny{2}};
        \node (E) at (3.5,-0.45) {\tiny{1}};
        \node (F) at (3.5,-1.65) {\tiny{2}};

        \draw (c) edge[->,thick,dashed] node[above, yshift=1pt] {a} (a)
              (c) edge[->,thick] node[above, yshift=1pt] {b} (b);
    \end{tikzpicture}}
    \\
\end{tabular}
\begin{tabular}{  p{2.345cm}  p{2.7cm}  p{3.0cm}  }

    \vspace{-1mm} del0(x:list) & \vspace{-1mm} set\_flag(x:list) & \vspace{-1mm} flag(x:list) \\

    \vspace{-2mm}
    \adjustbox{valign=t}{\begin{tikzpicture}
        \node (a) at (0.0,0) [root node] {x};

        \node (b) at (1.0,0) {$\Rightarrow$};
        
        \node (c) at (2.0,0) {$\emptyset$};
    \end{tikzpicture}}

    &

    \vspace{-2mm}
    \adjustbox{valign=t}{\begin{tikzpicture}
        \node (a) at (0.0,0) [root node] {x};

        \node (b) at (1.0,0) {$\Rightarrow$};

        \node (c) at (2.0,0) [root node, fill=gp2grey] {x};

        \node (A) at (0.0,-.45) {\tiny{1}};
        \node (C) at (2.0,-.45) {\tiny{1}};
    \end{tikzpicture}}

    &

    \vspace{-2mm}
    \adjustbox{valign=t}{\begin{tikzpicture}
        \node (a) at (0.0,0) [root node, fill=gp2grey] {x};

        \node (b) at (1.0,0) {$\Rightarrow$};

        \node (c) at (2.0,0) [root node, fill=gp2grey] {x};

        \node (A) at (0.0,-.45) {\tiny{1}};
        \node (C) at (2.0,-.45) {\tiny{1}};
    \end{tikzpicture}}
    \\
\end{tabular}
\end{allintypewriter}
\end{minipage}
}
\caption{The program \texttt{is-binary-dag.gp2}}
\label{fig:bin-dag-recognition-program-det}
\end{figure}

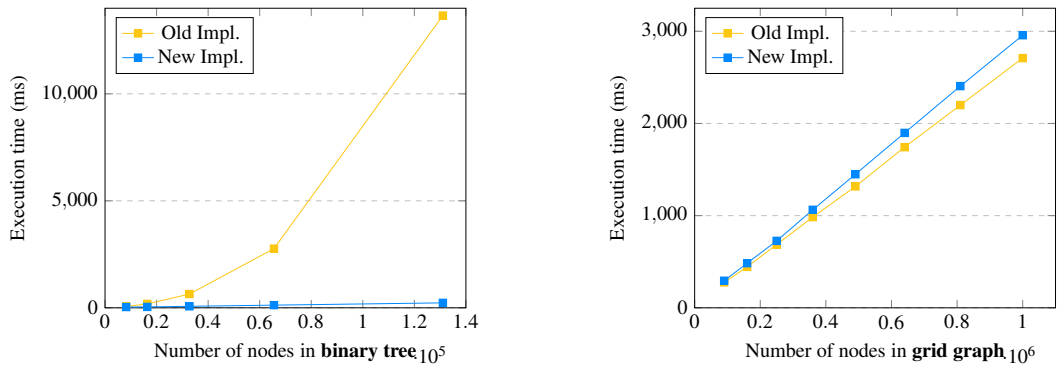
\begin{figure}[H]
\centering
{
\begin{subfigure}{.48\textwidth}
    \centering
    \pgfplotsset{scaled y ticks=false}
\begin{tikzpicture}[scale=0.7]
\begin{axis}[
xlabel={Number of nodes in \textbf{binary tree}},
ylabel={Execution time (ms)},
xmin=0, xmax=140000,
ymin=0, ymax=14000,
legend pos=north west,
ymajorgrids=true,
grid style=dashed,
yticklabel style={/pgf/number format/fixed},
]
\addplot[color=performanceYellow, mark=square*] 
coordinates {
    (8191,62.22)
    (16383,183.16)
    (32767,642.28)
    (65535,2765.78)
    (131071,13657.44)
};
\addplot[color=performanceBlue, mark=square*] 
coordinates {
    (8191,35.65)
    (16383,40.81)
    (32767,69.82)
    (65535,124.63)
    (131071,231.38)
};
\addlegendentry{Old Impl.}
\addlegendentry{New Impl.}
\end{axis}
\end{tikzpicture}
\end{subfigure}
\begin{subfigure}{.48\textwidth}
    \centering
    \pgfplotsset{scaled y ticks=false}
\begin{tikzpicture}[scale=0.7]
\begin{axis}[
xlabel={Number of nodes in \textbf{grid graph}},
ylabel={Execution time (ms)},
xmin=0, xmax=1100000,
ymin=0, ymax=3250,
legend pos=north west,
ymajorgrids=true,
grid style=dashed,
yticklabel style={/pgf/number format/fixed},
]
\addplot[color=performanceYellow, mark=square*] 
coordinates {
    (90000,275.13)
    (160000,444.56)
    (250000,685.72)
    (360000,983.92)
    (490000,1316.96)
    (640000,1742.48)
    (810000,2199.17)
    (1000000,2709.24)
};
\addplot[color=performanceBlue, mark=square*] 
coordinates {
    (90000,293.67)
    (160000,483.92)
    (250000,726.04)
    (360000,1063.11)
    (490000,1448.70)
    (640000,1897.87)
    (810000,2405.17)
    (1000000,2957.17)
};
\addlegendentry{Old Impl.}
\addlegendentry{New Impl.}
\end{axis}
\end{tikzpicture}
\end{subfigure}
}
\vspace{-0.5em}
\caption{Measured performance of \texttt{is-binary-dag.gp2}}
\label{fig:is-bin-dag-timing}
\end{figure}

The program implements a depth-first search for nodes without incoming edges, recording the \enquote{upward} path of the root node by dashed edges, combined with reduction steps that delete such nodes. The main loop either terminates with the empty graph, implying that the input graph was a binary DAG, or upon execution of the break statement when a node has been detected to which neither \texttt{up} nor one of the reduction rules is applicable. We observe that the original compiler produces a program that runs in quadratic time on many graph classes, including full binary trees and grid graphs. Our new compiler runs in linear time on such graphs (see Figure \ref{fig:is-bin-dag-timing}).

\begin{figure}[H]
\centering
\noindent
\fbox{\begin{minipage}{.8\textwidth}
\begin{allintypewriter}
Main = init; Reduce!; Unmark; if Check then fail

Reduce = \{prune0, prune1, push\}

Unmark = try unmark; try unmark

Check = \{two\_nodes, has\_loop\}

\setlength{\tabcolsep}{0.4cm}

\medskip
\smallskip

\begin{tabular}{ p{3.8cm} p{6cm} }
    
    init(x:list) & prune0(a,x,y:list) \\

    \begin{tikzpicture}
        \node (a) at (0.0,0) [gp2 node] {x};

        \node (b) at (1.0,0) {$\Rightarrow$};
        
        \node (c) at (2.0,0) [root node] {x};
        
        \node (A) at (0.0,-.45) {\tiny{1}};
        \node (C) at (2.0,-.45) {\tiny{1}};
    \end{tikzpicture}

    &

    \begin{tikzpicture}
        \node (a) at (0.0,0) [gp2 node] {x};
        \node (b) at (1.5,0) [root node] {y};
        
        \node (c) at (2.5,0) {$\Rightarrow$};
        
        \node (d) at (3.5,0) [root node] {x};
        
        \node (A) at (0.0,-.45) {\tiny{1}};
        \node (D) at (3.5,-.45) {\tiny{1}};
        
        \draw (a) edge[->,thick] node[above] {a} (b);
    \end{tikzpicture}
    \\
\end{tabular}

\vspace{-1.1em}

\begin{tabular}{ p{3.8cm} p{6cm} }
    
    unmark(x:list) & prune1(a,x,y:list) \\

    \begin{tikzpicture}
        \node (a) at (0.0,0) [gp2 node, fill=gp2grey] {x};

        \node (b) at (1.0,0) {$\Rightarrow$};
        
        \node (c) at (2.0,0) [gp2 node] {x};
        
        \node (A) at (0.0,-.45) {\tiny{1}};
        \node (C) at (2.0,-.45) {\tiny{1}};
    \end{tikzpicture}

    &

    \begin{tikzpicture}
        \node (a) at (0.0,0) [gp2 node, fill=gp2grey] {x};
        \node (b) at (1.5,0) [root node] {y};
        
        \node (c) at (2.5,0) {$\Rightarrow$};
        
        \node (d) at (3.5,0) [root node] {x};
        
        \node (A) at (0.0,-.45) {\tiny{1}};
        \node (D) at (3.5,-.45) {\tiny{1}};
        
        \draw (a) edge[->,thick] node[above] {a} (b);
    \end{tikzpicture}
    \\
\end{tabular}

\vspace{-1.1em}

\begin{tabular}{ p{10cm} }

    push(a,x,y:list) \\

    \begin{tikzpicture}
        \node (a) at (0.0,0)     [root node] {x};
        \node (b) at (1.5,0)     [gp2 node] {y};
        
        \node (c) at (2.5,0)     {$\Rightarrow$};
        
        \node (d) at (3.5,0)     [gp2 node, fill=gp2grey] {x};
        \node (e) at (5,0)       [root node] {y};
        
        \node (A) at (0.0,-.45) {\tiny{1}};
        \node (B) at (1.5,-.45) {\tiny{2}};
        \node (D) at (3.5,-.45) {\tiny{1}};
        \node (E) at (5.0,-.45) {\tiny{2}};
        
        \draw (a) edge[->,thick] node[above] {a} (b)
              (d) edge[->,thick] node[above] {a} (e);
    \end{tikzpicture}
    \\
\end{tabular}

\begin{tabular}{ p{3.2cm} p{7.8cm} }

    has\_loop(a,x:list) & two\_nodes(x,y:list) \\

    \begin{tikzpicture}
        \node (a) at (0.0,0) [gp2 node] {x};
        
        \node (b) at (1.0,0) {$\Rightarrow$};
        
        \node (c) at (2.0,0) [gp2 node] {x};
        
        \node (A) at (0.0,-.45) {\tiny{1}};
        \node (C) at (2.0,-.45) {\tiny{1}};
        
        \draw (a) edge[->,in=-30,out=-60,loop,thick] node[right, yshift=1.5pt] {a} (a)
              (c) edge[->,in=-30,out=-60,loop,thick] node[right, yshift=1.5pt] {a} (c);
    \end{tikzpicture}

    &

    \vspace{-10.8mm}
    \begin{tikzpicture}
        \node (a) at (0.0,0) [gp2 node] {x};
        \node (b) at (1.5,0) [gp2 node] {y};

        \node (c) at (2.5,0) {$\Rightarrow$};

        \node (d) at (3.5,0) [gp2 node] {x};
        \node (e) at (5.0,0) [gp2 node] {y};

        \node (A) at (0.0,-.45) {\tiny{1}};
        \node (B) at (1.5,-.45) {\tiny{2}};
        \node (D) at (3.5,-.45) {\tiny{1}};
        \node (E) at (5.0,-.45) {\tiny{2}};
    \end{tikzpicture}
    \\
\end{tabular}
\vspace{-5mm}
\end{allintypewriter}
\end{minipage}
}
\caption{GP\,2 Program \texttt{is-tree.gp2}}
\label{fig:is-tree-gp2}
\end{figure}

\begin{figure}[H]
\centering
\fbox{\begin{minipage}{.8\textwidth}
\begin{allintypewriter}
Main = link!

\setlength{\tabcolsep}{0.4cm}

\medskip
\smallskip

\begin{tabular}{ p{10.5cm} }
	
	link(a,b,x,y,z:list) \\
	
	\begin{tikzpicture}
        
		\node (a) at (0,0)       [gp2 node] {x};
		\node (b) at (1.25,0)    [gp2 node] {y};
		\node (c) at (2.5,0)     [gp2 node] {z};
		
		\node (d) at (3.375,0)   {$\Rightarrow$};
		
		\node (e) at (4.25,0)    [gp2 node] {x};
		\node (f) at (5.5,0)     [gp2 node] {y};
		\node (g) at (6.75,0)    [gp2 node] {z};
		
		\node (A) at (0,-.45)    {\tiny{1}};
		\node (B) at (1.25,-.45) {\tiny{2}};
		\node (C) at (2.5,-.45)  {\tiny{3}};
		\node (E) at (4.25,-.45) {\tiny{1}};
		\node (F) at (5.5,-.45)  {\tiny{2}};
		\node (G) at (6.75,-.45) {\tiny{3}};
		
		\draw (a) edge[->,thick] node[above] {a} (b)
		      (b) edge[->,thick] node[above] {b} (c)
		      (e) edge[->,thick] node[above] {a} (f)
		      (f) edge[->,thick] node[above] {b} (g)
		      (e) edge[->,thick,bend left=40] (g);
	\end{tikzpicture}
	\\
	
	\vspace{-1em}where not edge(1,3) \\
	
\end{tabular}
\end{allintypewriter}
\end{minipage}}
\caption{The program \texttt{transitive-closure.gp2}}
\label{fig:transitive-closure.gp2}
\end{figure}

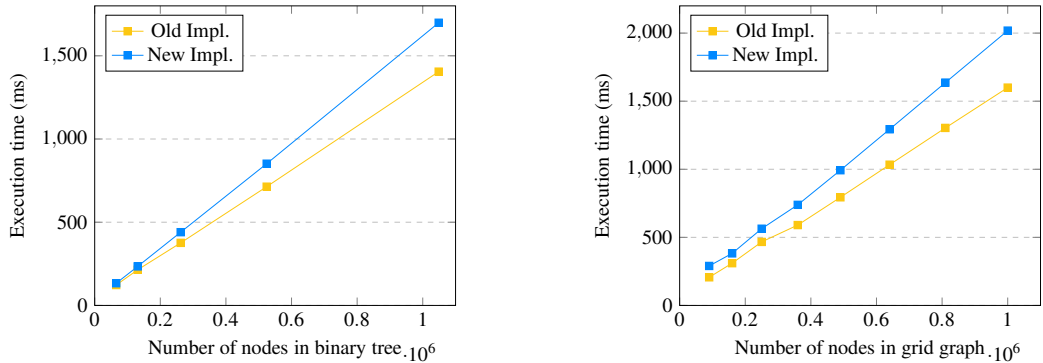
\begin{figure}[H]
\centering
{
\begin{subfigure}{.48\textwidth}
    \centering
    \pgfplotsset{scaled y ticks=false}
\begin{tikzpicture}[scale=0.7]
\begin{axis}[
xlabel={Number of nodes in binary tree},
ylabel={Execution time (ms)},
xmin=0, xmax=1100000,
ymin=0, ymax=1800,
legend pos=north west,
ymajorgrids=true,
grid style=dashed,
yticklabel style={/pgf/number format/fixed},
]
\addplot[color=performanceYellow, mark=square*] 
coordinates {
    (65535,122.52)
    (131071,214.20)
    (262143,375.95)
    (524287,713.55)
    (1048575,1404.75)
};
\addplot[color=performanceBlue, mark=square*] 
coordinates {
    (65535,133.40)
    (131071,234.47)
    (262143,440.12)
    (524287,851.67)
    (1048575,1698.62)
};
\addlegendentry{Old Impl.}
\addlegendentry{New Impl.}
\end{axis}
\end{tikzpicture}
\end{subfigure}
\begin{subfigure}{.48\textwidth}
    \centering
    \pgfplotsset{scaled y ticks=false}
\begin{tikzpicture}[scale=0.7]
\begin{axis}[
xlabel={Number of nodes in grid graph},
ylabel={Execution time (ms)},
xmin=0, xmax=1100000,
ymin=0, ymax=2200,
legend pos=north west,
ymajorgrids=true,
grid style=dashed,
yticklabel style={/pgf/number format/fixed},
]
\addplot[color=performanceYellow, mark=square*] 
coordinates {
    (90000,206.53)
    (160000,310.22)
    (250000,466.51)
    (360000,590.19)
    (490000,794.06)
    (640000,1033.59)
    (810000,1303.68)
    (1000000,1599.73)
};
\addplot[color=performanceBlue, mark=square*] 
coordinates {
    (90000,289.97)
    (160000,382.21)
    (250000,563.41)
    (360000,738.82)
    (490000,992.92)
    (640000,1293.69)
    (810000,1636.45)
    (1000000,2018.30)
};
\addlegendentry{Old Impl.}
\addlegendentry{New Impl.}
\end{axis}
\end{tikzpicture}
\end{subfigure}
}
\vspace{-0.5em}
\caption{Measured performance of \texttt{is-tree.gp2}}
\label{fig:is-tree-timing}
\end{figure}

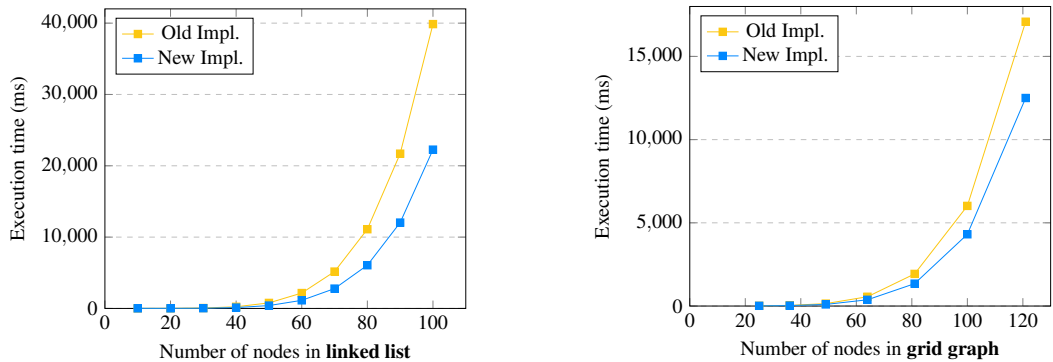
\begin{figure}[H]
\centering
{
\begin{subfigure}{.48\textwidth}
    \centering
    \pgfplotsset{scaled y ticks=false}
\begin{tikzpicture}[scale=0.7]
\begin{axis}[
xlabel={Number of nodes in \textbf{linked list}},
ylabel={Execution time (ms)},
xmin=0, xmax=110,
ymin=0, ymax=42000,
legend pos=north west,
ymajorgrids=true,
grid style=dashed,
yticklabel style={/pgf/number format/fixed},
]
\addplot[color=performanceYellow, mark=square*] 
coordinates {
    (10,11.50)
    (20,16.10)
    (30,56.63)
    (40,232.39)
    (50,777.02)
    (60,2158.89)
    (70,5165.75)
    (80,11106.40)
    (90,21699.39)
    (100,39864.74)
};
\addplot[color=performanceBlue, mark=square*] 
coordinates {
    (10,10.85)
    (20,12.97)
    (30,32.32)
    (40,119.44)
    (50,401.89)
    (60,1136.29)
    (70,2781.41)
    (80,6057.80)
    (90,12036.47)
    (100,22264.00)
};
\addlegendentry{Old Impl.}
\addlegendentry{New Impl.}
\end{axis}
\end{tikzpicture}
\end{subfigure}
\begin{subfigure}{.48\textwidth}
    \centering
    \pgfplotsset{scaled y ticks=false}
\begin{tikzpicture}[scale=0.7]
\begin{axis}[
xlabel={Number of nodes in \textbf{grid graph}},
ylabel={Execution time (ms)},
xmin=0, xmax=130,
ymin=0, ymax=18000,
legend pos=north west,
ymajorgrids=true,
grid style=dashed,
yticklabel style={/pgf/number format/fixed},
]
\addplot[color=performanceYellow, mark=square*] 
coordinates {
    (25,15.46)
    (36,38.96)
    (49,145.37)
    (64,558.13)
    (81,1930.91)
    (100,6017.70)
    (121,17083.91)
};
\addplot[color=performanceBlue, mark=square*] 
coordinates {
    (25,13.79)
    (36,27.41)
    (49,96.79)
    (64,375.40)
    (81,1345.13)
    (100,4311.57)
    (121,12500.58)
};
\addlegendentry{Old Impl.}
\addlegendentry{New Impl.}
\end{axis}
\end{tikzpicture}
\end{subfigure}
}
\vspace{-0.5em}
\caption{Measured performance of \texttt{transitive-closure.gp2}}
\label{fig:trans-closure-timing}
\end{figure}

Next, we look at a rooted tree reduction program by Campbell \cite{Campbell19a} (Figure \ref{fig:is-tree-gp2}) that was linear time on graphs of bounded degree in the previous implementation of the compiler. We confirm that it remains linear time (Figure \ref{fig:is-tree-timing}). Finally, we measure the performance of an unrooted program (Figure \ref{fig:transitive-closure.gp2}) which computes the transitive closure of a graph. The new compiler is superior on linked lists and grid graphs (Figure \ref{fig:trans-closure-timing}) due to the re-implementation of edge lists which speeds up the intense search for edges in the absence of root nodes.

\section{Conclusion and Future Work} \label{section:conclusion}

The new compiler improves the performance of some programs significantly while retaining the complexity class for others. The program \texttt{is-discrete.gp2} has been brought down to linear complexity due to our node list's capacity to skip holes in the underlying node array. Moreover, the new compiler outperforms the old when running \texttt{transitive-closure.gp2} by a significant constant factor. We attribute this to better cache usage; in the old compiler, only the indices of the first two edges are statically stored in the \texttt{Node} type, whereas the new compiler stores several more direct pointers to nodes statically in a node's big array. 

Some programs perform better or worse depending on the type of input graph. For example, the \texttt{is-binary-dag.gp2} program is accelerated from quadratic to linear time on binary trees and linked lists, but is slightly slower now on grid graphs. Such variance is to be expected with substantial implementation changes. Overall, the worst case drop in performance is by a constant factor, meaning no complexities were worsened in the observed test cases. Several programs perform similarly to before with a slightly enlarged constant, a byproduct of the memory operations that a linked-list data structure entails. Moreover, the new compiler resolves some fundamental bugs in the graph parser's implementation.

Future work should determine whether it makes sense to add lists of marked and unmarked nodes in order to find a marked or unmarked node in constant time. Ongoing research is also exploring what classes of graph algorithms can be implemented in linear time in GP\,2 using the current compiler. Bak and Plump showed that a depth-first search of graphs of bounded degree and with a bounded number of connected components can be performed in linear time \cite{Bak-Plump12a,Bak15a}, and also 2-colouring of such graphs. This paper shows that it is possible to execute a reduction algorithm for binary DAGs in linear time that doesn't limit the growth of the number of components.

\bibliographystyle{eptcs}
\bibliography{ms}
\end{document}